\documentclass{sig-alternate}

\usepackage[colorlinks,urlcolor=black,linkcolor=black,anchorcolor=black,citecolor=black,breaklinks=true,bookmarks=false,pdfstartview={FitH}]{hyperref}
\usepackage{breakurl}
\usepackage{color}
\usepackage[linesnumbered,ruled,vlined]{algorithm2e}
\usepackage{url}
\usepackage{amsmath}
\usepackage{multirow}
\usepackage{enumerate}
\usepackage{hyperref} 
\usepackage{booktabs}
\usepackage{colortbl}
\usepackage{threeparttable}
\usepackage{subfigure}

\usepackage{authblk} 

\usepackage{bigstrut}
\usepackage{makecell}

\usepackage{caption}

\makeatletter
\let\@copyrightspace\relax
\makeatother

\begin{document}

\title{A Scalable Pipelined Dataflow Accelerator for Object Region Proposals on FPGA Platform \titlenote{This work was supported in part by the National Natural Science Foundation of China (61602022, 61501013, 61571023, 61521091 and 1157040329), National Key Technology Program of China (No. 2017ZX01032101) and the 111 Talent Program B16001.}}

\author[1,3]{ Wenzhi Fu }
\author[1,3]{ Jianlei Yang }
\author[2,3]{ Pengcheng Dai }
\author[4]{ Yiran Chen }
\author[2,3]{ Weisheng Zhao }
\affil[1]{ School of Computer Science and Engineering, Beihang University, Beijing, 100191, China.}
\affil[2]{ School of Electronic and Information Engineering, Beihang University, Beijing, 100191, China.}
\affil[3]{ Fert Beijing Research Institute, BDBC, Beihang University, Beijing, 100191, China.}
\affil[4]{ Department of Electrical and Computer Engineering, Duke University, Durham, NC 27708, USA. \authorcr \{jianlei, weisheng.zhao\}@buaa.edu.cn}
\renewcommand\Authands{ and }

\maketitle

\begin{abstract}
Region proposal is critical for object detection while it usually poses a bottleneck in improving the computation efficiency on traditional control-flow architectures. We have observed region proposal tasks are potentially suitable for performing pipelined parallelism by exploiting dataflow driven acceleration. In this paper, a scalable pipelined dataflow accelerator is proposed for efficient region proposals on FPGA platform. The accelerator processes image data by a streaming manner with three sequential stages: resizing, kernel computing and sorting. First, Ping-Pong cache strategy is adopted for rotation loading in resize module to guarantee continuous output streaming. Then, a multiple pipelines architecture with tiered memory is utilized in kernel computing module to complete the main computation tasks. Finally, a bubble-pushing heap sort method is exploited in sorting module to find the top-$k$ largest candidates efficiently. Our design is implemented with high level synthesis on FPGA platforms, and experimental results on VOC2007 datasets show that it could achieve about 3.67X speedups than traditional desktop CPU platform and >250X energy efficiency improvement than embedded ARM platform.
\end{abstract}




\section{Introduction}\label{section:introduction}

Recent years, deep neural networks have made a great success in image classification tasks of single object \cite{krizhevsky2012imagenet}. However, it can not be directly applied to multi-object detection, which is much more practical in real-world applications \cite{zhang2015group,castanon2015efficient}. Object detection first decides the interested regions which potentially include objects (so-called region proposal) and then performs image classification on these proposed regions. Efficient real-time object detection is a prerequisite for many kinds of unmanned systems since the energy efficiency of intensive computation is critical for them. There have been many research works focused on hardware acceleration for image classification \cite{emer2016tutorial} but no one designed for region proposals, which have posed a bottleneck in efficient object detection.

Recently several computationally intensive approaches have shown a better performance \cite{Liu2016SSD}, however, it is not practical for them running on embedded platform with limited resources and restricted power consumption. On the other hand, binarized normed gradients (BING) method is proposed as an effective approach for region proposal generation \cite{cheng2014bing,zhang2017sequential,wei2016hcp:,zha2015exploiting}, which achieves state-of-the-art detection rate. However, even though there have already been several techniques to optimize the implementation of BING algorithm on CPU\cite{cheng2014bing}, due to the control-flow processing style of Von Neumann architecture, it is still difficult for BING to run efficiently especially in embedded platform.

In this work, a scalable dataflow accelerator is proposed for the BING algorithm who has been reformed to a dataflow-driven manner. With the memory hierarchy, this efficiency of the streaming processing can be highly improved which overcomes the shortages of the traditional platform. Furthermore, this accelerator has a good scalability to be scaled to a larger parallelism efficiently without the problem of synchronization.

The main contributions of this work are listed as following:

\begin{enumerate}[(1)]
\item The BING algorithm is reformed as a dataflow-driven manner to obtain significant benefits from dataflow processing.
\item A dataflow architecture is proposed for the BING algorithm, which generate a continues input stream then process it in a streaming manner to fully deployed the locality and minimized the intermediate data amount.
\end{enumerate}


\section{Region Proposal with BING}\label{section:preliminaries}

The aim of region proposal is to find the interested regions which potentially contain objects with minimized windows, which is important in multi-object detection tasks. Previous works \cite{cheng2014bing,zhang2017sequential} have shown that a simple $8 \times 8$ feature by computing the normed gradients could be adopted as an efficient region proposal. After binarizing the normed gradients feature, BING could achieve efficient objectness estimation. The original image is first resized to different sizes with different preset resizing ratios, therefore, the region proposal candidates with different resolution in the original image can be represented with $8 \times 8$ window uniformly in the resized images. Then, the SVM (Support Vector Machine) stage \uppercase\expandafter{\romannumeral1} is adopted to evaluate the confidence for each region proposal. Following that, the NMS (non-maximum suppression) stage is utilized to reduce the overlap among the proposals. After that, top-$n$ largest candidates are selected from the region proposals corresponding to each resized image and the SVM stage \uppercase\expandafter{\romannumeral2} is performed to evaluate the confidence among all of the resized images. Finally, the top-$k$ largest candidates could be obtained as final proposals by a sorting stage. Since the BING algorithm has few branch or jump operations, it could be abstracted as a data-driven algorithm which is ideal to perform dataflow driven acceleration.

\section{Proposed Dataflow Accelerator}\label{section:accelerator}

\subsection{Accelerator Framework}

\begin{figure}
    \centering
    \subfigure[Accelerator framework]{
        \label{figure:fig2:architecture}
        \includegraphics[width=0.45\textwidth]{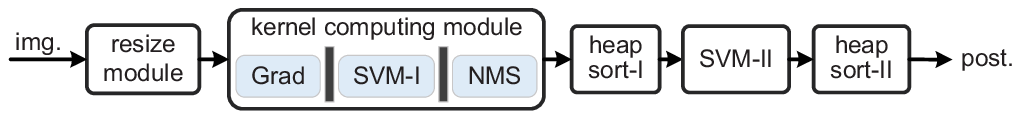}}
    \subfigure[Pipelined kernel computing driven by resized images]{
        \label{figure:fig2:kernel}
        \includegraphics[width=0.4\textwidth]{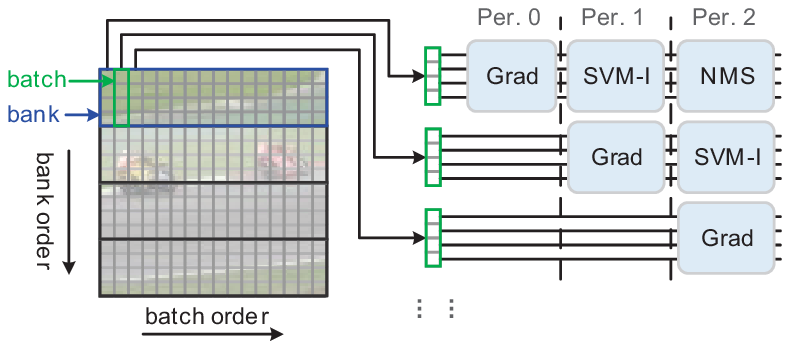}}
    \caption{Demonstration of our proposed accelerator.}
    \label{figure:fig2}
\end{figure}

 Corresponding to the computation process of BING algorithm, the framework of the proposed dataflow accelerator shown in Fig. \ref{figure:fig2:architecture} can also be divided into resizing module, kernel computing module and sorting module. The function of each module is same as the corresponding computation in the algorithm but can be completed in a different manner. As shown in Fig. \ref{figure:fig2:architecture}, the resizing module first resize the original image with preset ratios then partition the resized image into a series of batch, which represent for four neighbor pixels vertically. The following kernel computing module works in the granularity of batch and can be divided into three stages: CalcGrad, SVM and NMS operations, which can process the continuously batch streaming with a parallel pipelines architecture and exports a stream of candidates. The sorting module is deigned to obtain the top-$k$ or top-$n$ candidates by performing bubble-pushing heap sort strategy to satisfy the throughput requirements. Finally, the region proposals could be obtained after post processing.
 Without loss of generality, the kernel computing module is demonstrated by four pipelines in this paper which could be extended as more pipelines as following. And the bubble-pushing heap sort model is very similar to \cite{zabolotny2011dual} which is too trivial to be listed here.

\subsection{Resizing Module}

\begin{figure}
    \centering
    \includegraphics[width=0.48\textwidth]{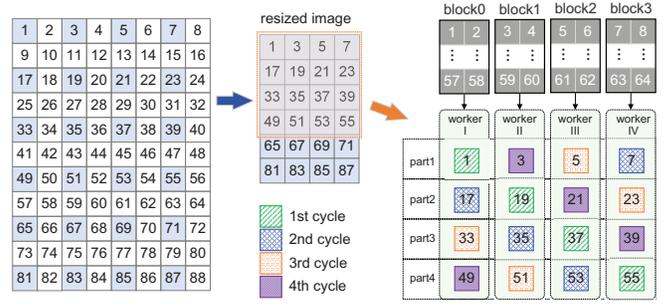}
    \caption{Illustration of resizing module.}
    \label{figure:fig3}
\end{figure}

\begin{figure}
    \centering
    \includegraphics[width=0.48\textwidth]{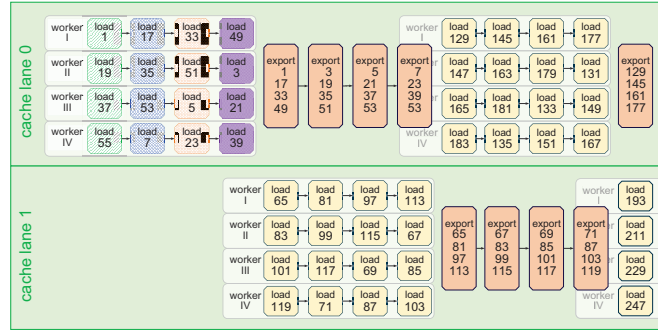}
    \caption{Continuous batch streaming with Ping-Pong cache.}
    \label{figure:fig4}
\end{figure}

 A naive resizing approach is carried out in \cite{srivastava2017accelerating} but it cannot satisfy our requirements for streaming processing. The kernel computing module requires a continuous batch streaming output from resizing module to make sure the multiple pipelines are fully loaded. In this work, the original image is partitioned into four blocks uniformly as shown in Fig. \ref{figure:fig3}. Only one port of the configured BRAMs is assigned for each block while two dual-port or four single-port BRAM are required for processing each image. The pixels from four blocks are fetched in parallel as processed by four workers. Since the fetched pixels from different blocks are discontinuous, a Ping-Pong cache, which consists of two cache lanes, is adopted here for buffering. Meanwhile, the cache is also partitioned as four parts which correspond to the four BRAM ports. The fetching procedure loads the pixels in different blocks by a rotation style and feeds them to different part of cache. As shown in Fig. \ref{figure:fig4}, two groups of workers on two cache lanes could alternately provide continuous batch streaming with Ping-Pong cache strategy.

\subsection{Kernel Computing Module}

The kernel computing module includes CalcGrad, SVM-I and NMS stage. First, we define a distance in RGB color space between pixel $P_a$ and $P_b$ as
\begin{displaymath}
D\left( {{P_a},{P_b}} \right) = \mathop {\max }\limits_{q \in \left\{ {R,G,B} \right\}} \left| {{P_a}\left( q \right) - {P_b}\left( q \right)} \right|
\end{displaymath} The gradients on vertical direction and horizon direction are defined as ${I_x}\left( {i,j} \right)$ and ${I_y}\left( {i,j} \right)$, respectively, where $i$ and $j$ represents the pixel location. They could be obtained by calculating the normed gradient between neighbor pixels
\begin{displaymath}
{I_x}\left( {i,j} \right) = D\left( {{P_{\left( {i - 1,j} \right)}},{P_{\left( {i + 1,j} \right)}}} \right) \\
\end{displaymath}
\begin{displaymath}
{I_y}\left( {i,j} \right) = D\left( {{P_{\left( {i,j - 1} \right)}},{P_{\left( {i,j + 1} \right)}}} \right)
\end{displaymath} and the gradient of each pixel $G\left( {i,j} \right)$ could be calculated by
\begin{displaymath}
G\left( {i,j} \right) = \min \left\{ {{I_x}\left( {i,j} \right) + {I_y}\left( {i,j} \right),255} \right\}
\end{displaymath} The obtained normed gradient of each pixel is reformulated as a two-dimension array.

In the SVM-I stage, the normed gradients ${G_{8 \times 8}}$ of every $8 \times 8$ window are formed by the gradients ${G_{1 \times 8}}$ of each row and reshaped as a $64$-dimension feature with a row-wise manner. Then the trained SVM weights ${W_{SVM}}$ are adopted to perform the classification and determine the evaluation scores of each window
\begin{displaymath}
s = {G_{8 \times 8}} \cdot {W_{SVM}}
\end{displaymath} And all the scores $s$ compose a two-dimensional array $S$. During the NMS stage, the max score ${\max _{5 \times 5}}$ for each ${5 \times 5}$ block of $S$ is determined by finding the max score ${\max _{1 \times 5}}$ for each row first and then maximum of them. For all ${5 \times 5}$ blocks, only the window corresponding to ${\max _{5 \times 5}}$ will be selected for windows sequence output.

\begin{figure}
    \centering
    \includegraphics[width=0.45\textwidth]{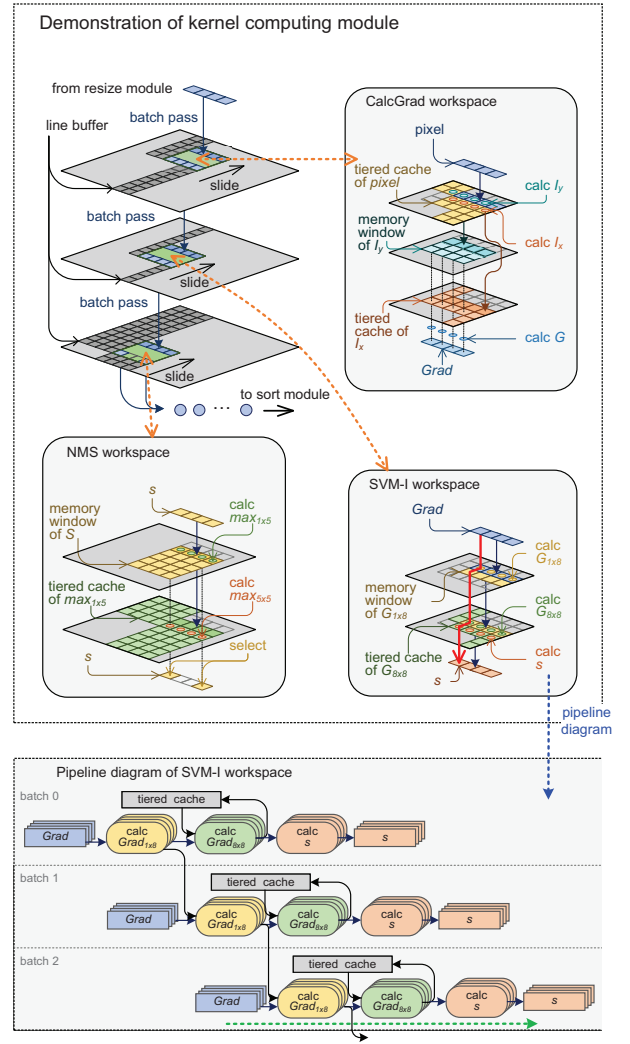}
    \caption{Demonstration of kernel computing module with the diagram of SVM-I pipeline.}
    \label{figure:fig6}
\end{figure}

A rough approach to perform CalcGrad, SVM-I and NMS tasks usually requires a temporal two-dimensional array for intermediate data buffering which can result in waste of resources and computation cycles. In our design, these stages are reformulated and delivered on multiple pipelines for streaming processing as shown in Fig. \ref{figure:fig6}. Each of the three stages has its own workspace with its own pipelines which performs operations in a streaming manner, and can be connected serially for processing batch streaming continuously. Meanwhile, the data locality in each workspace is exploited by the tiered cache, which is built with memory window and line buffer\cite{vallina2012implementing}, by caching all the required data for batch pass synchronization.

Since the pipelines behave similarly, the SVM-I stage is taken for pipelines illustration as an example without loss of generality. As shown in Fig. \ref{figure:fig6}, the input batch streaming can be processed continuously by the pipelines and consequently generate a streaming output for the following stage. During the processing of each workspace, the data are continuously loaded from tiered cache for the calculation of ${G_{8 \times 8}}$, while the calculation results are restored to the tiered cache for the next operation simultaneously.

At the end of the kernel computing module, the NMS operation usually results in non-continuous output streaming. In this work, a FIFO structure is adopted as streaming buffer to make sure the pipelines run smoothly which could improve the total efficiency of the proposed accelerator.

\section{Experimental Results}\label{section:result}

\subsection{Experiment Setup}

The proposed accelerator is implemented by C language and synthesized with Xilinx Vivado HLS 2017 \cite{vivado2017xilinx} on two target chips: Artix-7 (low voltage version) @ 3.3MHz for always-on \& low power application, and Kintex UltraScale+ @ 100MHz for real-time \& high performance application. A carefully quantization strategy is adopted to specify various bit-width for different data storage purpose. The synthesized resources utilizations are illustrated in Table \ref{tab:fpga}.
The power consumption and system latency is obtained by C-RTL co-simulation in Vivado. The VOC2007 datasets \cite{voc2007dataset} are adopted to evaluate the quality of proposed windows on the metrics by detection rate (DR v.s. \#WIN), and mean average best overlap (MABO v.s. \#WIN), where \#WIN is the number of given proposals. The metric DR v.s. \#WIN means the detection rate (DR) for the given \#WIN proposals \cite{cheng2014bing}. And MABO v.s. \#WIN means the mean average best overlap for the given \#WIN proposals \cite{zhang2017sequential}. Hence, a larger DR or MABO value means a better proposal quality.

\begin{figure}
    \centering
    \includegraphics[width=0.45\textwidth]{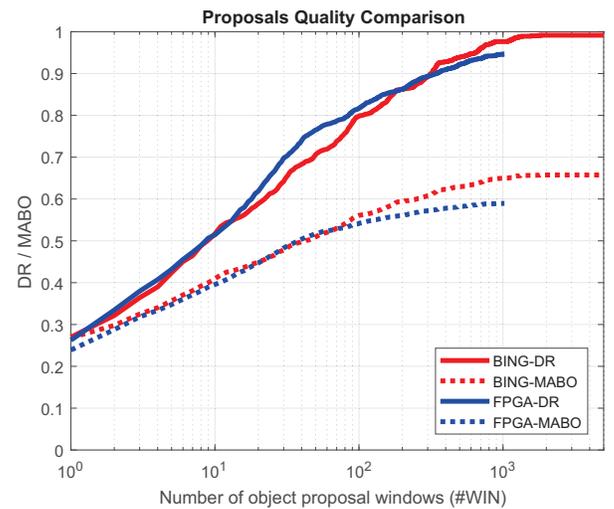}
    \caption{Quality evaluation of windows proposals by comparing BING \cite{cheng2014bing} and our proposed FPGA accelerator.}
    \label{figure:fig10}
\end{figure}

\subsection{Performance Evaluation}

A defined IoU (intersection-over-union) parameter in the previous works \cite{zhang2017sequential} is also adopted in our evaluation. It is a scoring function to measure the affinity of two bounding boxes, defined by the intersection area of two bounding boxes divided by their union. The DR and MABO metrics are measured by varying the IoU overlap threshold which is set as $0.4$ as default for correct detection.

\begin{table}[htb]
\centering
\caption{The FPGA resources utilizations comparison between two target devices: Artix-7 and Kintex UltraScale+.}
\label{tab:fpga}
\begin{threeparttable}
\begin{tabular}{c|c|c|c|c}
    \cline{1-5}
    \multirow{2}{*}{Resources}  & \multicolumn{2}{c|}{\makecell[c]{Artix-7 Low Volt.\tnote{\dag} \\  @ 3.3MHz}}  & \multicolumn{2}{c}{\makecell[c]{Kintex UltraScale+\tnote{$\ast$} \\ @ 100MHz}} \bigstrut \\
    \cline{2-5}
     & Available & Utilized & Available & Utilized  \bigstrut \\
    \cline{1-5}
     LUT  &  63400 & 54453 &  162720 &   56504   \bigstrut \\
     LUT-RAM  &  19000  &   4166  &  99840   & 3157   \bigstrut \\
     FF  &  126800  &  48611   &   325440  &  50079   \bigstrut \\
     BRAM  &  135  &  135   &   360  & 146   \bigstrut \\
     DSP  &  240  &   25  &   1368  & 25   \bigstrut \\
     BUF-G  &  -  &   -  &  256   &  8   \bigstrut \\
    \cline{1-5}
\end{tabular}
\begin{tablenotes}
\item [\dag] Targeted on Artix-7 (low voltage) xc7a100tlftg256-2L.
\item [$\ast$] Targeted on Kintex UltraScale+ xcku3p-ffva676-3-e.
\end{tablenotes}
\end{threeparttable}
\end{table}

The detection accuracy of our proposed accelerator is compared with BING \cite{cheng2014bing} as shown in Fig. \ref{figure:fig10}. The BING algorithm generates 5000 object windows to evaluate the accuracy. However, we have observed it only achieves less than $3\%$ accuracy improvement compared with only $1000$ object windows generated. Hence, only $1000$ object windows are proposed in our design when considering the significantly increasing requirements on hardware resources. For evaluating $1000$ proposals, the detection rate of our proposed approach (denoted as FPGA-DR) is about $94.72\%$ while BING is about $97.63\%$.

\begin{table}[htb]
\centering
\caption{Speedup and power efficiency compared with Intel i7 and ARM platforms.}
\label{tab:comparison}
\begin{threeparttable}
\begin{tabular}{c|c|c|c|c}
\hline
    & \multicolumn{2}{l|}{Kintex UltraScale+} & \multicolumn{2}{l}{Artix-7 Low Volt.}
    \\ \cline{2-5}
    & Speedup & \makecell[l]{Power \\ efficiency } & Speedup & \makecell[l]{Power \\ efficiency }
    \\ \hline
    Intel i7 & 3.67X & \textgreater{}220X  & 0.12X   & 66X
    \\ \hline
    ARM A53 & 68X & \textgreater{}250X  & 2.2X    & \textgreater{}60X
    \\ \hline
\end{tabular}
\end{threeparttable}
\end{table}


\begin{table}[htb]
\centering
\caption{Performance evaluation between targeted on Artix-7 and Kintex UltraScale+, where ${P_{tot}}$ is the total power consumption which includes static power and dynamic power consumption, and ${P_{dyn}}$ is the dynamic power consumption.}
\label{tab:performance}
\begin{tabular}{c|c|c|c|c|c}
    \cline{1-6}
    \multicolumn{3}{c|}{\makecell[c]{Artix-7 Low Volt. \\ @ 3.3MHz}}  & \multicolumn{3}{c}{\makecell[c]{Kintex UltraScale+ \\ @ 100MHz}} \bigstrut \\
    \cline{1-6}
      ${P_{tot}}$ & ${P_{dyn}}$ & Speed   & ${P_{tot}}$ & ${P_{dyn}}$ & Speed     \bigstrut \\
    \cline{1-6}
      $97mW$	  & $15mW$      & $35fps$ & $821mW$     & $350mW$     & $1100fps$ \bigstrut \\
    \cline{1-6}
\end{tabular}
\end{table}

The implemented FPGA accelerators are compared with two conventional CPU platforms. The BING algorithm is well-optimized and could achieve a proposal speed by $300fps$ on Intel i7-3940XM CPU (TDP: 55W) platform with multithreaded programming and subword parallelism techniques \cite{cheng2014bing}. We also evaluate BING on a Raspberry-Pi 3B platform with 64-bit ARM A53 processor, which could achieve $16fps$ speed and 3W$\sim$4W \cite{raspi2018power} power consumption. However, the proposed accelerator on Kintex UltraScale+ target could achieve a proposal speed by $1100fps$ while the power consumption is only $821mW$ when running at 100MHz, which is applicable for real-time processing of multi-camera sensor fusion applications. Hence, it could achieve about 3.67X speedups and >220X energy efficiency improvement compared with BING on Intel i7 CPU. Furthermore, the proposed accelerator targeted on low voltage Artix-7 could achieve $35fps$ with an extremely low power consumption $97mW$ when running at 3.3MHz, which is attractive for ultra-low power applications with always-on working mode whose speed is also sufficient for most of the applications. The detailed comparison results of our proposed FPGA accelerators against the two CPU platforms are shown in Table \ref{tab:comparison}.


\section{Conclusions}\label{section:conclusion}

Efficient region proposal generation is a critical task for object detection. A scalable dataflow accelerator is proposed in this paper for efficient region proposals on FPGA platform. The proposal procedures in BING algorithm are reformulated as a dataflow driven manner and implemented on multiple pipelines architecture. All of the modules in the accelerator are organized by streaming processing mechanisms so that these streaming data could guarantee the pipelines are fully loaded. Furthermore, a tiered cache system is utilized to improve the bandwidth of data synchronization between different processing stages. Evaluations on VOC2007 datasets show that our proposed accelerator achieves a very large scale of speedups and energy efficiency improvement with a little detection rate degradation.



\end{document}